%% file: 00-ArXiv-Transitions.tex
\documentclass{article}

\usepackage{arXiv}

\usepackage[utf8]{inputenc} 
\usepackage[T1]{fontenc}    
\usepackage{hyperref}       
\usepackage{url}            
\usepackage{booktabs}       
\usepackage{amsfonts}       
\usepackage{nicefrac}       
\usepackage{microtype}      
\usepackage{lipsum}		
\usepackage{graphicx}
\usepackage{doi}

\usepackage[numbers,sort&compress]{natbib}
\title{On transitions in water wave propagation through consolidated to broken sea ice covers}

\author{
Jordan P.A.~Pitt \\
	University of Adelaide, Australia \\
	\texttt{jordan.pitt@adelaide.edu.au} 
\And
Luke G.~Bennetts \\
	University of Adelaide, Australia \\
	\texttt{luke.bennetts@adelaide.edu.au} 
}

\date{}

\renewcommand{\shorttitle}{Wave propagation through consolidated to broken sea ice covers}

\hypersetup{
pdftitle={On transitions in water wave propagation through consolidated to broken sea ice covers},
pdfsubject={q-bio.NC, q-bio.QM},
pdfauthor={Jordan~P.A.~Pitt, Luke G.~Bennetts},
pdfkeywords={First keyword, Second keyword, More},
}

\input{arxiv-luke.tex}

\begin{document}
\maketitle
\begin{abstract}
A theoretical model is used to study water waves propagating into and through a region containing thin floating ice, for ice covers transitioning from consolidated (large floe sizes) to fully broken (small floe sizes). 
The degree of breaking is simulated by a mean floe length.
It is shown that there are deterministic limits for consolidated and fully broken ice covers where the wave fields do not depend on the particular realisation of the ice cover for a given mean floe length.
The consolidated ice limit is consistent with classic flexural-gravity wave theory, and the fully broken limit is well modelled by Bloch waves in a periodic ice cover.
In the transition between the limits, the wave field depends on the ice cover realisation, as multiple wave scattering is a dominant process.
The effects of the ice cover on the wave field are quantified using a wavelength, attenuation rate, and a transferred amplitude (measuring the amplitude drop at the ice edge).
It is shown that as the ice cover breaks up (mean floe size gets smaller), the wavelength and amplitude drop decreases (transferred amplitude increases) and the attenuation rate increases. The results provide a new interpretation of field observations.
\end{abstract}


\section{Introduction}

Understanding and modelling ocean surface wave propagation through the outer band of the sea ice covered ocean (known as the marginal ice zone) is important to predict the responses of Arctic and Antarctic sea ice covers to climate change \cite{squire2011past,bennetts2023closing} and calving of icebergs \cite{holdsworth1978iceberg,massom2018antarctic}, and to improve operational forecasts in high latitude oceans \cite{squire2013better,aksenov2017future}.
It has been the subject of intense research activity over the past decade or so, involving contributions from geophysicists, engineers, mathematicians and more, on observations, theory, and numerical and physical modelling \cite{thomson2018overview,Squire-2020,golden2020modeling,bennetts2022theory}.
Global wave models are being extended to generate predictions of waves in the ice covered oceans \cite{doble2013wave,rogers2021estimates}, and sea ice models are integrating wave impacts on the ice cover \cite{Bennetts-2017,williams2017wave,roach2018emergent,bateson2020impact}, thereby empowering studies of coupled wave--ice feedbacks \cite{dumont2022marginal,horvat2022floes,thomson2022wave}.

Wave--ice interaction models are generally based on the coupled processes of wave-induced ice breakup and wave attenuation due to the ice cover \cite{dumont2011wave,Williams-2013a,Williams-2013b}.
Ocean waves impose flexural strains on consolidated ice covers that break them into an unconsolidated collections of ice floes \cite{Kohout-2016,herman2021sizes,dumas2023aerial}, i.e., 
breakup of a given ice cover depends on both wavelengths and wave amplitudes.
Unconsolidated/broken ice covers are more susceptible to drift in response to winds and currents \cite{alberello2020drift}, and to melt in the summer \cite{horvat2022floes}.
The ice cover attenuates waves over distance, so that, beyond a certain distance (that depends on the properties of the incident waves and the prevailing ice cover), the imposed strains become too small to break the ice \cite{Robin-1963}. 
Therefore, wave attenuation models are needed to predict the extent of wave-induced ice breakup and the resulting changes in wave and ice dynamics \cite{herman2022granular,horvat2022floes,mokus2022wave,pitt2022model}. 

Observations of waves in the ice covered ocean show that amplitudes of the spectral components tend to attenuate at an exponential rate over distance \cite{Squire-1980,wadhams1988attenuation}, and
the attenuation coefficient (the rate of exponential attenuation) has a power-law dependence on wave frequency \cite{Meylan-2014,Meylan-2018}, such that higher frequency components (shorter waves) are attenuated over shorter distances. 
Wave scattering models, in which wave interactions with individual floes are resolved, reproduce the exponential attenuation and low-pass filter properties, and compare well with historical data for the regime in which wavelengths are comparable to floe lengths \cite{kohout2008elastic,Bennetts-2012}. 
Attenuation due to scattering becomes negligible in the regimes where wavelengths are either much smaller than floe lengths (consolidated ice covers) or much greater than floe lengths (unconsolidated/fully broken ice covers) \cite{Meylan-2021}. 
A plethora of continuum sea ice models have been developed for the consolidated and fully broken regimes, in which attenuation results from damping due to anelastic flexure, viscoelasticity, parameterised turbulence, etc.~\cite{shen2022wave}. 
At present, only the model in which damping results from drag pressure proportional to the particle velocity at the wave--ice interface 
\cite{Williams-2013a}
(often referred to as the Robinson--Palmer damping model \cite{Robinson-1990}) gives a power-law dependence comparable to observations \cite{Meylan-2018}.
This damping is often combined with scattering to construct a model capable of reasonable wave attenuation predictions for consolidated to fully broken ice covers
\cite{vaughan2009decay,Bennetts-2012a,Meylan-2021,mokus2022wave}.

Wave attenuation observations are usually not accompanied by detailed information on the ice cover, such as ice thickness and floe sizes \cite{Alberello-2021}.
The large variation in observed attenuation coefficients has been attributed to an underlying dependence on the ice properties \cite{montiel2022physical,stopa2018strong}.
Collins~et~al.~\cite{Collins-2015} observed waves from a vessel in the ice covered ocean, 
in initially consolidated ice conditions, for which the ice blocked incident waves from reaching the vessel, and after a wave event had broken the ice, for which the incident waves were largely unimpeded from reaching the vessel.
Ardhuin~et~al.~\cite{ardhuin2020ice} observed waves using drifting buoys and interpreted consolidated and broken ice states using satellite images. 
They also noted a sharp increase in wave activity following ice breakup.
Both Collins~et~al.~\cite{Collins-2015} and Ardhuin~et~al.~\cite{ardhuin2020ice} interpreted their observations as wave attenuation decreasing following break up of consolidated ice covers. 


Physical models have been conducted in wave--ice tanks of wave propagation through an initially consolidated floating ice cover that is subsequently broken by waves into floes much shorter than the wavelengths, where waves were measured at a series of distances into the ice covered water and cameras monitored the ice breakup \cite{Dolatshah-2018,Passerotti-2022}.
They reveal sharp drops in wave amplitude across the edges of consolidated ice covers, which models of wave interactions with landfast ice predict \cite{fox1994oblique}, but are not accounted for in attenuation models.
The physical models also indicate that attenuation rates are greater in broken ice covers than consolidated ice covers.
Therefore, following breakup of a consolidated ice cover, waves will appear stronger if the observation location is close to the ice edge (due to the reduced ice edge amplitude drop) or weaker if the observation location is far from the ice edge (due to the increased attenuation over distance).

In this study, we use a theoretical model of wave--ice interactions to investigate both the ice edge amplitude drop and the attenuation coefficient, for ice conditions from consolidated (large floes) to fully broken (small floes).
We show that the model predicts the ice edge amplitude drop effect becomes weaker and the attenuation coefficient increases as the ice cover transitions from consolidated to fully broken, in qualitative agreement with the physical models, and providing an alternative interpretation of observations \cite{Collins-2015,ardhuin2020ice}.
For completeness, we also study the change in wavelength through the transition.
We use homogenization to derive semi-analytical expressions for the quantities of interest in the fully broken regime, 
and compare the dispersion curves with those of flexural-gravity and mass-loaded waves. 

\section{Preliminaries}

\begin{figure}
\centering
\includegraphics[width=\textwidth]{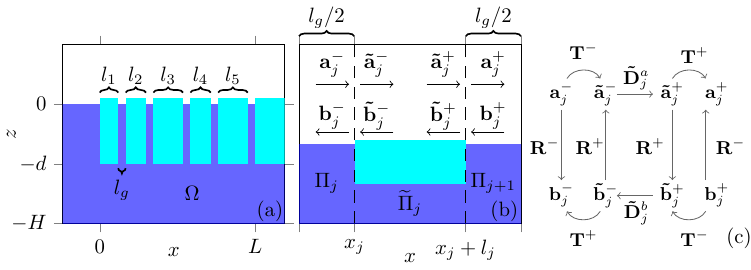} 
\caption{(a)~Schematic of an example ice cover with $N=5$. (b)~The wave mode amplitude representation for the $j^{th}$ floe. 
(c)~The reflection, transmission and propagation relationships between the amplitudes vectors for the $j^{th}$ floe. }
\label{Fig:ModelSchematic}
\end{figure}

\subsection{Problem Description}

Let $x$ and $z$ denote the horizontal and vertical coordinates, respectively. 
Water occupies a domain $\Omega$ (Figure~\ref{Fig:ModelSchematic}a) from its unloaded (open water), undisturbed free-surface ($z=0$) to the impermeable seafloor ($z=-H$) and extends infinitely out to the left ($x\to{}-\infty$) and right ($x\to{}\infty$). 
The water has density $\rho$ and undisturbed depth $H$, such that $H=20\lambda\gg\lambda$ (where $\lambda$ is the wavelength) to approximate deep water conditions. 
The water is incompressible, inviscid and irrotational, with the location of its free surface, $z=\hat{\eta}(x,t)$, a small perturbation from its location at rest ($z=0$). 
Thus, linear potential flow theory is applied to model the water motion, using the velocity potential $\hat{\phi}(x,z,t)$, for which the water velocity field is $(u,v) = \nabla \hat{\phi}$, where $\nabla\equiv(\partial/\partial{}x,\partial/\partial{}z)$.

A floating ice cover occupies $x>0$. It is composed of ice with thickness $h$, density $\rho'$ and, hence, Archimedean draft $d = (\rho'/\rho) h $. The ice is modelled as a linear Kirchoff-Love plate, with Young’s modulus $E$, Poisson's ratio $\nu$, and viscous (Robinson-Palmer) damping defined by a damping coefficient $\gamma$. 
As a Kirchoff-Love plate, the ice deformation can be described using the deformation of its bottom surface , i.e., the one-dimensional displacement of the ice-water interface. 

The ice cover in $0<x<L$ is broken into $N$ floes with lengths $l_{1},\dots,l_{N}$ (from left to right). 
The remaining ice ($x>L$) is consolidated, i.e., a semi-infinite floe.
The floes are separated from each other by open water gaps of constant length $l_g$. 
Therefore, the interval of broken ice cover ($0<x<L$) contains alternating sections of open and ice covered water.
The water covered by the $j^{th}$ floe is denoted $\widetilde{\Pi}_{j}$ and the open water to the left of the $j^{th}$ floe is denoted ${\Pi}_{j}$ (Figure~\ref{Fig:ModelSchematic}b). Thus, ${\Pi}_{1}$
is the open water occupying $x<0$ and $\widetilde{\Pi}_{N +1}$ is the consolidated ice cover ($x > L$). 

Time-harmonic motions are assumed at angular frequency $\omega$, such that the surface displacement and the velocity potential can be expressed as, respectively,
\begin{equation}\label{eqn:Wave_BaseForm}
 \hat{\eta}(x,t) = \Re \left\lbrace {\eta}(x) e^{-i\omega t} \right\rbrace \quad\text{and}\quad
  \hat{\phi}(x,z,t) = \Re \left\lbrace \frac{g}{i \omega}{\phi}(x,z) e^{-i\omega t}\right\rbrace,
\end{equation} 
where the reduced surface elevation ${\eta}(x) $ and velocity potential ${\phi}(x,z)$ are complex valued. 
The governing equations are expressed, in terms of the velocity potential, as
\begin{subequations}\label{eqn:Gov_Eqn}
\begin{align}
\partial^2_x {\phi} + \partial^2_z {\phi} &= 0 &\text{for } x,z \in \Omega, \label{eqn:NonDim_Eqs_1}\\
\partial_z {\phi} &= 0 &\text{for } x\in\Omega \text{ and } z = -H, \\
\partial_z {\phi} - \sigma {\phi} &= 0 &\text{for } x \in {\Pi}_{j} \text{ and } z=0, \label{eqn:NonDim_Eqs_2} \\
\left( F\partial_x^4 - S \sigma - i \Gamma\sqrt{\sigma} + 1 \right)\partial_z {\phi} - \sigma {\phi} &= 0 &\text{for } x \in \widetilde{\Pi}_{j} \text{ and } z=-d ,\label{eqn:NonDim_Eqs_3} \\
\partial_x {\phi} &= 0 & \text{for } x=x_j \wedge x=x_{j}+l_j \text{ and } z \in [0,-d], \label{eqn:NonDim_Eqs_4}\\
\partial_x^2 \partial_z {\phi}
=\partial_x^3 \partial_z {\phi} &= 0 &\text{for } x=x_j \wedge x=x_{j}+l_j \text{ and } z \in [0,-d], \label{eqn:NonDim_Eqs_5}
\end{align}
\end{subequations}
where (\ref{eqn:Gov_Eqn}c--f) hold for $j=1,2,\ldots,N,N+1$.
Equation~(\ref{eqn:Gov_Eqn}a) is Laplace's equation, (\ref{eqn:Gov_Eqn}b) is the impermeable seafloor condition and (\ref{eqn:Gov_Eqn}c) is the open water free-surface condition, where $\sigma = \omega^2 /g$ is a frequency parameter.
Equation~(\ref{eqn:Gov_Eqn}d) is the ice-coupled surface condition, where 
\begin{equation}
 F = \frac{E h^3}{12g\rho(1 - \nu^2)}, 
 \quad
 S = \frac{ \rho' h}{\rho} 
 \quad\text{and}\quad 
 \Gamma = \frac{\gamma}{\sqrt{g}\rho}
\end{equation} 
represent, respectively, scaled versions of the flexural rigidity, mass per unit area and damping rate of the ice.
Equation~(\ref{eqn:Gov_Eqn}e) is no water flow through the submerged portions of the ice edge, and (\ref{eqn:Gov_Eqn}f) are free edge conditions at the ice edges (zero bending moment and shear stress).
The surface elevation is recovered from the velocity potential as $\eta(x)=\phi(x,0)$ in $\Pi_{j}$
and $\eta(x)=\phi(x,-d)$ in $\widetilde{\Pi}_{j}$ ($j=1,2,\ldots,N,N+1$).

Motions of the water--ice system are forced by unit amplitude waves incident to the ice cover from the open water region ${\Pi}_{1}$. 
Thus, the governing equations (\ref{eqn:Gov_Eqn}) are closed by the radiation conditions
\begin{subequations}\label{eqn:rad_conds}
\begin{align}
\phi(x,z) &\sim \left( e^{i\kappa_{1}x} + Re^{-i\kappa_{1}x}\right)\frac{\cosh(\kappa_{1}(z+H))}{\cosh(\kappa_{1}H)} &\text{as } x\to -\infty,\\[8pt] 
\phi(x,z) &\sim Te^{ik_{1}x}\frac{\cosh(k_{1}(z+H))}{\cosh(k_{1}(H-d))} &\text{as } x\to \infty,
\end{align}
\end{subequations}
where the wavenumbers $\kappa_{1}$ and $k_{1}$ are defined in \S\ref{sec:soln}.

\subsection{Solution Method}\label{sec:soln}
For each $\Pi_{j}$ and $\widetilde{\Pi}_{j}$ ($j=1,2,\ldots,N,N+1$),
${\phi}(x,z)$ is represented by a linear superposition of wave modes, which have the form
\begin{equation}\label{eqn:Wave_BaseForm}
  e^{\pm i\kappa x}\frac{\cosh(\kappa(z+H))}{\cosh(\kappa H)} \quad \text{for} \quad x \in {\Pi}_{j} \quad\text{and}\quad
  e^{\pm i k x}\frac{\cosh(k(z+H))}{\cosh(k(H-d))} \quad \text{for} \quad x \in \widetilde{\Pi}_{j}. 
\end{equation}
The wavenumbers $\kappa$ and $k$ satisfy dispersion relations produced by the different boundary conditions \eqref{eqn:NonDim_Eqs_2} and \eqref{eqn:NonDim_Eqs_3}, which are
\begin{subequations}\label{eqn:Disp_Rel}
\begin{align}\label{eqn:Disp_Rel_1}
 \kappa \tanh(\kappa H) &= \sigma, 
 \\ \label{eqn:Disp_Rel_2}
 \left(Fk^4 - S \sigma - i \Gamma \sqrt{\sigma} + 1 \right) k \tanh(k (H-d)) &= \sigma. 
\end{align}
\end{subequations}
The open-water dispersion relation \eqref{eqn:Disp_Rel_1} has one positive real root, $\kappa=\kappa_1$, and infinitely many purely imaginary roots, with $\kappa=\kappa_2,\dots,\kappa_{M}$ being the $M-1$ smallest solutions on the positive imaginary axis, ordered by increasing magnitude. 
The real wavenumber corresponds to a lossless propagating wave in the open water and the purely imaginary roots correspond to evanescent waves that decay exponentially at a rate that depends on the magnitude of the wavenumber. 
For $\Gamma =0$ (no damping), the ice-coupled dispersion relation 
\eqref{eqn:Disp_Rel_2} also has one positive real root, $k=k_1$, corresponding to a propagating flexural-gravity wave, and infinitely many purely imaginary roots (corresponding to evanescent waves) with $k=k_2,\dots,k_{M}$ being the $M-1$ smallest solutions on the positive imaginary axis, ordered by increasing magnitude. 
The addition of damping to \eqref{eqn:Disp_Rel_2} ($\Gamma \neq0$) perturbs the non-damped case, so that $k_1$ is complex and, thus, the propagating flexural-gravity wave decays, while $k_2,\dots,k_{M}$ are no longer purely imaginary. 
For the parameters used in this study, the real parts of $k_1,\dots,k_M$ were less than three orders of magnitude smaller than their imaginary parts and, thus, these modes effectively behave as purely evanescent waves. The ice-coupled dispersion relation \eqref{eqn:Disp_Rel_2} has two additional complex roots, denoted $k_{M+1}$ and $k_{M+2}$, 
which correspond to vertical modes that depend linearly on the modes supported by the propagating and evanescent roots \cite{Bennetts-2007}.




Linearly superposing the wave modes, the velocity potentials can be approximated as
\begin{subequations}\label{eqn:Phi_SumForm}
\begin{align}
 &{\phi}(x,z) =\sum^{M}_{m=1} \left(a_{m,j} e^{i \kappa_m (x - x_{j})} + b_{m,j} e^{-i \kappa_m (x - x_{j})} \right) \frac{\cosh(\kappa_m(z+H))}{\cosh(\kappa_m H)} \quad \text{for} \quad x \in {\Pi}_{j}, \label{eqn:Phi_SumForm1} \\
 &{\phi}(x,z) = \sum^{M+2}_{m=1} \left(\tilde{a}_{m,j} e^{i k_m (x - x_{j})} + \tilde{b}_{m,j} e^{-i k_m (x - x_{j})} \right) \frac{\cosh(k_m(z+H))}{\cosh(k_m(H-d))} \quad \text{for} \quad x \in \widetilde{\Pi}_{j},
\end{align}
\end{subequations}
where $a_{m,j}$ and $b_{m,j}$ are the amplitudes of the left and rightward propagating wave modes in the open water at the left floe edge $x_j$. In the ice-covered water, the amplitudes of the leftward and rightward propagating wave modes at the left floe edge $x_j$ are $\tilde{a}_{m,j}$ and $\tilde{b}_{m,j}$. For the results presented, $M=200$ is used. 
The amplitudes in each region are combined into vectors $\mathbf{a}_j= [a_{1,j},\dots,a_{M,j}]$ and $\mathbf{b}_j= [b_{1,j},\dots,b_{M,j}]$ in the open water and $\mathbf{\tilde{a}}_j= [\tilde{a}_{1,j},\dots,\tilde{a}_{M+2,j}]$ and $\mathbf{\tilde{b}}_j= [\tilde{b}_{1,j},\dots,\tilde{b}_{M+2,j}]$ in the ice-covered water (Figure~\ref{Fig:ModelSchematic}b). 
The amplitude vectors $\mathbf{{a}}^-_j$, $\mathbf{{b}}^-_j$, $\mathbf{\tilde{a}}^-_j$, $\mathbf{\tilde{b}}^-_j$ are the amplitudes of the modes in the open and ice covered water at the left floe edge ($x_j$) and $\mathbf{{a}}^+_{j}$, $\mathbf{{b}}^+_{j}$, $\mathbf{\tilde{a}}^+_{j}$, $\mathbf{\tilde{b}}^+_{j}$ are the amplitudes of the modes in the open and ice covered water at the right floe edge ($x_{j}+l_j$).


Scattering of waves by the floe edges is described by reflection and transmission matrices, produced by matching variational forms of the continuity of the water pressure and the horizontal velocity across the floe edges and enforcing the free-edge conditions \eqref{eqn:NonDim_Eqs_4} and \eqref{eqn:NonDim_Eqs_5} \cite{Bennetts-2007}. The action of the reflection and transmission matrices for the $j^{th}$ floe is depicted in Figure~\ref{Fig:ModelSchematic}c. For example, consider $\mathbf{a}^{-}_{j}$, which represents rightward propagating wave modes coming into the left edge of the floe ($x_{j}$). 
These waves modes are partially reflected to generate $\mathbf{R}^- \mathbf{a}^{-}_{j}$ leftward propagating wave modes in the open water (contributing to $\mathbf{{b}}^{-}_{j} $) and partially transmitted to generate $\mathbf{T}^- \mathbf{a}^{-}_{j}$ rightward propagating wave modes in the floe covered region (contributing to $\mathbf{\tilde{a}}^{-}_{j}$). 
Similarly, the partial transmission and reflection of the wave modes coming into the left floe edge ($x_j$) from the right ($\mathbf{\tilde{b}}^{-}_{j}$) completes the expressions for scattering of incoming amplitudes, which produce the outgoing amplitudes, such that
\begin{align}
 \mathbf{\tilde{a}}^{-}_{j} = \mathbf{T}^- \mathbf{{a}}^{-}_{j} + \mathbf{R}^+\mathbf{\tilde{b}}^{-}_{j} \quad\text{and}\quad
 \mathbf{b}^{-}_{j} = \mathbf{R}^- \mathbf{a}^{-}_{j} + \mathbf{T}^+\mathbf{\tilde{b}}^{-}_{j}.
  \label{eqn:RefTrans_ScatEdge}
 \end{align}
Along the $j^{\text{th}}$ floe ($\widetilde{\Pi}_{j}$), as the floe properties are uniform, the wave modes simply propagate and decay in their direction of travel (over the floe length $l_j$), so that the amplitudes at the floe edges are related like so
\begin{subequations}\label{eqn:Prop_D_Defn}
\begin{align}
 \mathbf{\tilde{a}}^{+}_{j} &= \text{diag}\left\lbrace \left[ e^{i k_1 l_{j}}, \dots, e^{i k_{M+2} l_{j} }\right] \right\rbrace \mathbf{\tilde{a}}^{-}_{j} \equiv \mathbf{\tilde{D}}^{a}_j\mathbf{\tilde{a}}^{-}_{j} , \\[10pt]
 \mathbf{\tilde{b}}^{-}_{j} &= \text{diag}\left\lbrace \left[ e^{-i k_1 l_{j}}, \dots, e^{-i k_{M+2} l_{j} }\right] \right\rbrace \mathbf{\tilde{b}}^{+}_{j} \equiv \mathbf{\tilde{D}}^{b}_j\mathbf{\tilde{b}}^{+}_{j} .
 \end{align}
\end{subequations}
The transfer from ice to water is similar to \eqref{eqn:RefTrans_ScatEdge}, with the appropriate reflection ($\mathbf{R}^\pm $) and transmission matrices ($\mathbf{T}^\pm $). 
The wave modes propagating and/or decaying across the open water gaps produce similar expressions to \eqref{eqn:Prop_D_Defn}, with the appropriate wavenumbers ($\kappa$) and open water gap length ($l_g$). 

The propagation of the incident wave into the ice cover is generated by determining $\mathbf{a}^{\pm}_{j}$, $\mathbf{b}^{\pm}_{j}$, $\mathbf{\tilde{a}}^{\pm}_{j}$ and $\mathbf{\tilde{b}}^{\pm}_{j}$ for $j=1,2,\ldots,N,N+1$, using the reflection, transmission and  propagation/decay matrices, forced by the incident amplitudes of wave modes from the left of the ice cover ($\mathbf{a}^{-}_{1} = [1,0,\dots,0]$) and the (zero) incident wave modes from the right ($\mathbf{\tilde{b}}^{-}_{N+1} = [0,\dots,0]$). The multiple scattering problem is solved recursively, including the effects of scattering from all the $\Pi_{j}$ and $\widetilde{\Pi}_{j}$ regions \cite{Bennetts-2007} to obtain the solution to \eqref{eqn:Gov_Eqn} and \eqref{eqn:rad_conds}. 
To employ this method, the open-water regions $\Pi_j$ are numerically modelled as sections with extremely thin ice, $h_{\text{thin}} = 10^{-8}\, \text{m}$, with matching Kirchoff-Love plate parameters $\rho'$, $E$ and $\nu$ to the ice cover in $\widetilde{\Pi}_j$ and zero damping ($\Gamma=0$). 
The wavenumbers in the open water are obtained by solving \eqref{eqn:Disp_Rel_2} with $h_{\text{thin}}$, which is numerically equivalent to \eqref{eqn:Disp_Rel_1} for the regime of parameter values investigated in this paper, but with the addition of the two `numerical' wavenumbers ($\kappa_{M+1}$, $\kappa_{M+2}$). 
The wave modes associated with the two `numerical' wavenumbers are included in the open water \eqref{eqn:Phi_SumForm1}, resulting in square reflection ($\mathbf{R}^\pm$) and transmission matrices ($\mathbf{T}^\pm$). Similar to \cite{bennetts2022modeling}, the ice thickness $h_{\text{thin}} = 10^{-8}$\,m was set thin enough so that $\kappa_{1},\dots,\kappa_{M}$ and the corresponding entries in the reflection and transmission matrices well approximate those for the open water problem described above, and, thus, the numerical $\Pi_{j}$ sections behave as open water.

\subsection{Qualitative Results: Consolidated to Broken Ice Covers}

\begin{figure}
\centering
\includegraphics[width=\textwidth]{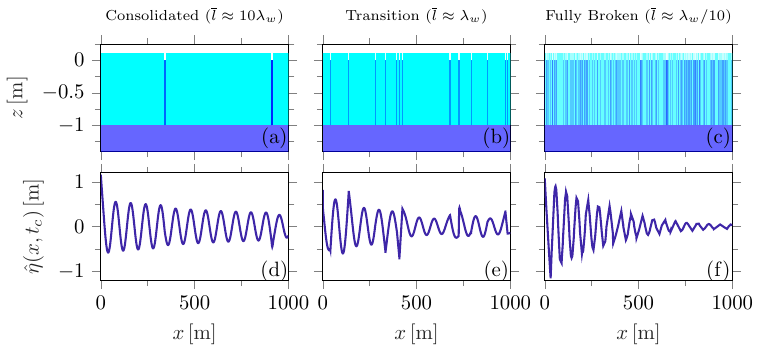} 
\caption{ (a,b,c)~Geometry of example ice covers with mean floe length (a)~$\bar{l} =625 \, \text{m} \approx 10\lambda_w $, (b)~$\bar{l} = 62.5 \, \text{m} \approx \lambda_w$, (c)~$\bar{l} =6.28 \, \text{m} \approx \lambda_w/10$. (d,e,f)~Numerical solutions for the water surface displacement $\hat{\eta}(x,t_c)$ for the ice covers depicted in (a), (b) and (c), respectively. }
\label{Fig:Disp_Only}
\end{figure}

Dimensional values of $h$, $g$, $E$, $\rho$, $\rho'$ and $\nu$ are chosen to give $F = 10^4$ and $S=1$, which are comparable to 
the values $F \approx 5.4\times 10^4$ and $S\approx0.897$ for sea ice covers \cite{Weeks-2010}. 
The non-dimensional damping parameter is set as $\Gamma = 0.1$, which is much greater than values typically used in wave--ice models, e.g., $\Gamma\approx0.0042$ \cite{Williams-2013a}. 
The larger value ensures that the wave energy reaching the consolidated ice at $x=L = 2500\,\text{m}$ is small and, thus, reflections from the consolidated ice ($x>L$) are negligible. 

The floe lengths are
generated by 'randfixedsum' \cite{RandSum}, which, for a given $N$, 
ensures that the total length of the broken ice cover section is fixed ($L$) and that all floes have lengths satisfying $10^{-5}\,\text{m}\le l_j\le L$. The lower limit on floe sizes is enforced to maintain numerical accuracy. The generated distribution of floe sizes is a uniform sampling of the space where $\sum_{j=1}^N (l_j + l_g) = L$, which resembles a power-law distribution where short floes are  more likely than long floes. Increasing $N$ makes short floes even more likely than long floes, increasing the apparent power of the generated floe size distribution.
The floes are separated by negligible length water gaps (set at $l_g = 10^{-12}\,\text{m}$ for numerical purposes), ensuring that the ice concentration is $\approx{}100\%$. Consequently, the average floe length is $\bar{l} \approx L / N$. 

The angular frequency is set as $\omega=1 \, \text{rad}\,\text{s}^{-1}$, which corresponds to the open water wavenumber $\kappa_1 = 0.1\,\text{rad}\,\text{m}^{-1}$ and wavelength $\lambda_w = 2\pi/ \kappa_1 = 20 \pi \approx 63 \,\text{m}$. 
The ice-coupled wavenumber is $k_1 \approx 0.078 + 9\times10^{-4}i \,\text{rad}\,\text{m}^{-1}$, resulting in a wavelength $\lambda_i = 2\pi/ \Re\left\lbrace k_1\right\rbrace \approx 80 \,\text{m} $ and attenuation rate $\alpha_i = \Im\left\lbrace k_1\right\rbrace \approx 9\times10^{-4} \,\text{m}^{-1}$. 
Three randomly generated realisations of the ice cover are considered, with different mean floe lengths:
$\bar{l} = 625 \, \text{m} \approx 10 \lambda_w $ ($N=4$; Figure~\ref{Fig:Disp_Only}a);
$\bar{l} = 62.5 \, \text{m} \approx \lambda_w$ ($N=40$; Figure~\ref{Fig:Disp_Only}b); 
and $\bar{l} =6.28\, \text{m} \approx \lambda_w/10$ ($N=398$; Figure~\ref{Fig:Disp_Only}c).
Numerical solutions are obtained for the example realisations, for $\bar{l} \approx 10 \lambda_w $ (Figure~\ref{Fig:Disp_Only}d),
$\bar{l} \approx \lambda_w$ (Figure~\ref{Fig:Disp_Only}e), and
 $\bar{l} \approx \lambda_w/10$ (Figure~\ref{Fig:Disp_Only}f).
The displacements of the free surface at time $t=t_c$, when $\hat{\eta}(0,t) = |\eta(0)|$, are shown to remove phase differences when comparing the waves in the different ice covers.

The wave fields for the three example ice covers demonstrate the effects of moving from an approximately consolidated ice regime, where floes are long compared to water wavelength (Figure~\ref{Fig:Disp_Only}d), 
to a fully broken ice regime, 
where floes lengths are small compared to water wavelengths (Figure~\ref{Fig:Disp_Only}f),
through a transition regime, where floes lengths are comparable to water wavelengths (Figure~\ref{Fig:Disp_Only}e). 
In the consolidated ice regime, the wave field resembles a decaying regular wave (which is consistent with the flexural-gravity wavenumber $k_1$ in intervals of unbroken ice cover). 
In the transition regime, wave scattering is dominant, which results in discontinuities in the wave field at the floe edges and an irregular wave field. 
In the fully broken ice regime, the floes have little individual effect on the passing waves and the wave field resembles a decaying regular wave, as if the ice cover were a homogeneous medium. 
However, the wave field in the fully broken ice regime is different to the one in the consolidated ice regime, and also in the open water (which produces no attenuation).


\section{Quantifying Waves-in-ice Properties}


\subsection{Wavelength, Attenuation Rate and Transferred Amplitude}

\begin{figure}
\centering
\includegraphics[width=\textwidth]{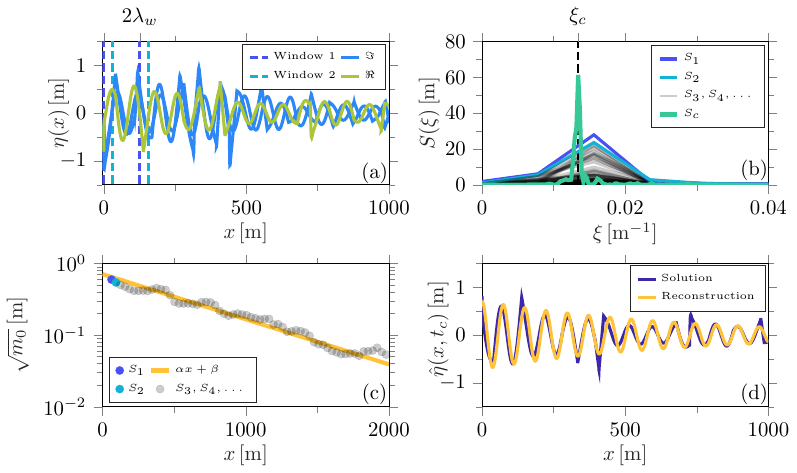} 
\caption{(a)~The real and imaginary parts of $\eta(x)$ for the ice cover with $\bar{l} \approx \lambda_w$ with the first two overlapping windows to generate the windowed spectra shown. (b)~The spectra over the whole ice cover $S_c$ and the windowed spectra $S_1,S_2,\dots$ for $\eta(x)$ as a function of spatial frequency $\xi$. (c)~Log-lin plot of the square root of zeroth moment for each windowed spectra compared to the fitted regression ($\alpha x + \beta$). (d)~Comparison of $\hat{\eta}(x,t_c)$ and the reconstruction based on the measured wavelength, attenuation rate and amplitude.}
\label{Fig:Spec_Method}
\end{figure}

To compare the properties of the observed wave fields, a regular wave reconstruction of the form
\begin{equation}
 \eta(x)\approx{}A \exp\left( i \left(\dfrac{2\pi}{\lambda} + i \alpha \right) x\right)\quad \text{for} \quad 0<x<L
  \label{eqn:Simple_Wave}
\end{equation}
is employed. 
A spectral method is applied to the numerical solution for the surface displacement, $\eta(x)$, to 
measure the dominant wavelength $\lambda$, overall attenuation rate $\alpha$ and the transferred wave amplitude into the ice cover $A$. 
Figure~\ref{Fig:Spec_Method} demonstrates 
the steps of the spectral method.
The ice cover from the transition regime is used ($\bar{l}\approx\lambda_{w}$; Figure~\ref{Fig:Disp_Only}b), 
for which scattering by the ice edges creates an irregular wave field over $0<x<L$
(Figure~\ref{Fig:Spec_Method}a).
This makes the wave fields in the transition regime the most challenging to reconstruct.

To calculate $\lambda$, the spectrum of $\eta(x)$ over the complete interval of broken ice ($0<x<L$), 
denoted $S_c(\xi)$, is used (Figure~\ref{Fig:Spec_Method}b).
The spectrum $S_c(\xi)$ describes the distribution of power of the wave field $\eta(x)$ over the spatial frequencies $\xi$. The location of the $S_c$-maximum, denoted $\xi_c$, indicates the dominant spatial frequency of the wave field and, thus, the dominant wavelength $\lambda = {2\pi}/{ \xi_c}$.
Due to wave attenuation, the `dominant' wavelength is biased to locations near the ice edge ($x=0$) where wave energy is largest. 
Removing the trend in attenuation did not significantly change the presented results. 

To measure $A$ and $\alpha$, $\eta(x)$ over $0<x<L$ is converted into windows of length $2 \lambda_w$ with $75\%$ overlaps (Figure~\ref{Fig:Spec_Method}a). 
For each window, the associated spectra are calculated, providing $S_1(\xi)$, $S_2(\xi)$ and so on (Figure~\ref{Fig:Spec_Method}b). 
The total energy in each window is given by the zeroth moment $m_0 = \int_{0}^\infty S(\xi) d\xi$. 
The average amplitude ($|\eta(x)|$) in the window can be estimated by $\sqrt{m_0}$. 
Figure~\ref{Fig:Spec_Method}c shows $\sqrt{m_0}$ for each window against the location of the window midpoint, and demonstrates 
a log-linear trend in $\sqrt{m_0}$ due to attenuation (caused by both scattering and dissipation).
To describe the observed trend in $\sqrt{m_0}$, the log-linear regression
\begin{equation}
 \ln(\sqrt{m_0}) \sim \alpha x + \beta
\end{equation}
is fitted, which produces $\alpha$ as the slope, and $\beta$ as the intercept. 
The transferred amplitude into the ice cover is the fitted $\sqrt{m_0}$ at the ice edge ($x=0$), so that $A = e^\beta $. 
Figure~\ref{Fig:Spec_Method}(d) shows the reconstructed wave field, using the measured values of $\lambda$, $\alpha$ and $A$, against the original wave field.


%

\subsection{Ensemble Averaging}
\begin{figure}
\centering
\includegraphics[width=1\textwidth]{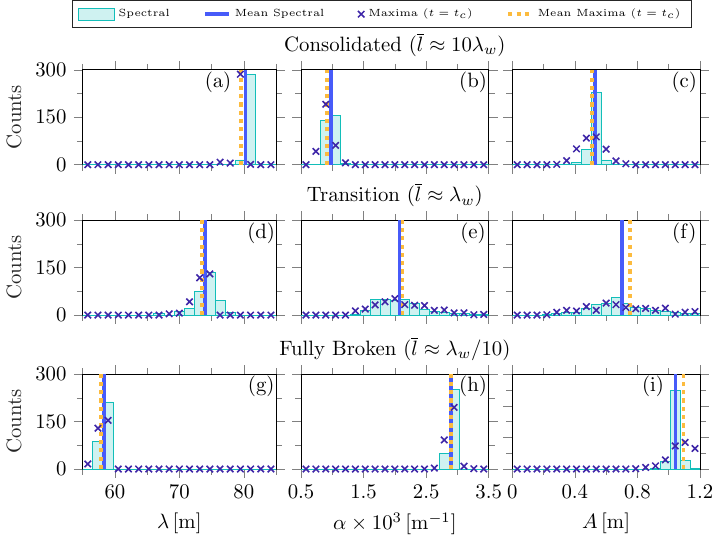} 
\caption{Histograms of measured (a,d,g)~$\lambda$, (b,e,h)~$\alpha$  and (c,f,i)~$A$,  
for ensembles of size $300$ as the mean floe lengths changes from (a,b,c)~$\bar{l} \approx10\lambda_w$ , (d,e,f)~$\bar{l} \approx\lambda_w$ and (g,h,i)~$\bar{l} \approx\lambda_w/10$.}
\label{Fig:Ensemble_Distro}
\end{figure}

The example ice covers in the consolidated, transition and fully broken regimes provide single instances of the ensemble of possible ice covers produced by random variations in floe length. 
The distributions of $\lambda$, $\alpha$ and $A$ are investigated
for three mean floe lengths ($10\lambda_w, \lambda_w$ and $\lambda_w/10$), 
 by randomly generating members of the ensemble and measuring the reconstructed wave properties. 
For each mean floe length, $300$ members are used, which was found to adequately represent the distributions. Figure~\ref{Fig:Ensemble_Distro} shows histograms for the reconstruction measures and, thus, the distribution of the ensembles for the ice cover regimes.

For consolidated ice (Figures~\ref{Fig:Ensemble_Distro}a--c), the ensemble has a consistent wavelength and attenuation rate, 
i.e., the wave field is insensitive to the particular realisation of the ice cover.
The amplitude $A$ has the greatest variability due to the different scattering near the front of the ice cover produced by the random floe lengths there. 
For the transition regime (Figures~\ref{Fig:Ensemble_Distro}d--f), the variances of all measures of the ensemble increases as wave scattering is a dominant process in this regime.
Consequently, ice covers in the transition regime are sensitive to random variations in the floe lengths and the ice cover cannot be described as a homogeneous medium. 
For fully broken ice covers (Figures~\ref{Fig:Ensemble_Distro}g--i), the variance decreases substantially, producing a consistent wavelength, attenuation rate and transferred amplitude for the entire ensemble, despite the somewhat irregular appearance of its members 
(Figure~\ref{Fig:Disp_Only}f). These results suggest that when the ice cover is sufficiently broken, it behaves as a homogeneous medium with a single $\lambda$, $\alpha$ and $A$ that are insensitive to the random variations in the floe lengths. 

The distributions of $\lambda$, $\alpha$ and $A$ given by the spectral method are compared against those given by a method using the maxima of $\hat{\eta}(x,t_c)$ (see Appendix~A). 
In general, the maxima method agrees with the spectral method for members of the ensembles and, thus, produces similar ensemble distributions. 
However, for fully broken ice the maxima method consistently overestimates $A$ compared to the spectral method and the original wave field. 
The spectral method is preferred for this study as it produces a better representation of the overall behaviour of the wave field. This is because the spectral method uses the average amplitude over the windows to determine 
$\alpha$ and $A$ and the spectrum over the complete broken ice interval to determine $\lambda$
instead of maxima location and values which are more sensitive to local behaviour.

\subsection{Results vs Mean Floe Length}


\begin{figure}
\centering
\includegraphics[width=\textwidth]{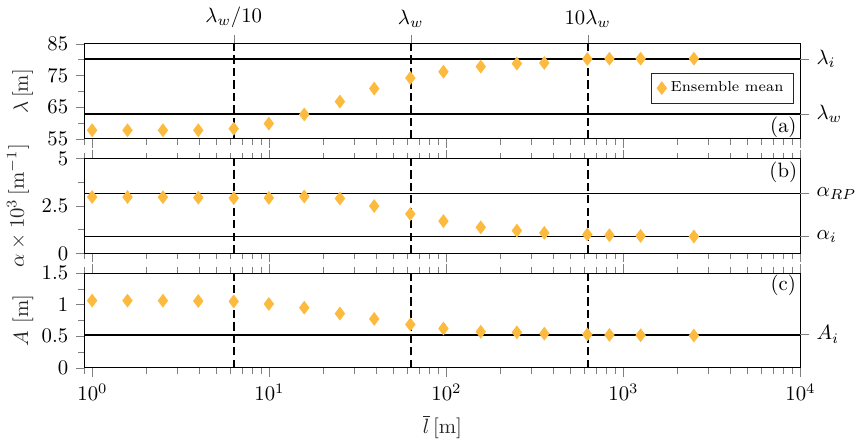} 
\caption{Ensemble means ($300$ members each) for (a)~$\lambda$, (b)~$\alpha$ and (c)~$A$, versus mean floe length. }
\label{Fig:FloeLen_Mean}
\end{figure}

Figure~\ref{Fig:FloeLen_Mean} shows the ensemble means of wavelength, attenuation rate and transferred amplitude versus mean floe length. 
The results for a continuous ice cover, $\lambda_{i}\approx{}80$\,m$>\lambda_{w}\approx{}63$\,m, $\alpha_{i}\approx 9\times{}10^{-4}$\,m$^{-1}$ and $A_{i}\approx{}0.5$\,m, which are obtained by solving the problem for $L=0$, are superimposed on their respective panels.
The wavelengths, attenuation rates and transferred amplitudes from the ensembles coincide with those of the continuous ice cover for $\bar{l} > 10 \lambda_w$.
This limiting behaviour corresponds to the consolidated ice regime, for which the wave fields for individual realisations are indistinguishable from one another (Figure~\ref{Fig:Ensemble_Distro}a--c).

For $\lambda_{w}/10<\bar{l}<10\lambda_{w}$, 
the wavelength increases with increasing floe length, $\bar{l}$, 
while the attenuation rate and transferred amplitude both decrease. 
This corresponds to the transition regime, where wave scattering causes the wave fields to depend on the individual realisations of the ice cover and the 
$\lambda$-, $\alpha$- and $A$-values for individual realisations at a given $\bar{l}$ are distributed around the ensemble means (Figure~\ref{Fig:Ensemble_Distro}d--f).

All three quantities reach plateaus for $\bar{l} < \lambda_w/10$, 
for which the wavelength $\lambda\approx{}58$\,m$<\lambda_{w}$ and the ice edge amplitude drop is negligible ($A\approx{}1$\,m).
The attenuation rate $\alpha\approx{}\alpha_{RP}=3\times 10^{-3}$\,m$^{-1}$, where $\alpha_{RP}$ is the imaginary part of the wavenumber solution to \eqref{eqn:Disp_Rel_2} when only the Robinson-Palmer damping term is retained and the deep-water approximation is used ($\tanh{k(H-d)} \approx 1$), such that
\begin{equation}
 k_{RP} = \frac{ \sigma}{1 - i\Gamma \sqrt{\sigma}} = {\frac{ \sigma}{1 + \Gamma^2 \sigma}}+ i \underbrace{\frac{ \sigma \sqrt{\sigma} \Gamma}{1 + \Gamma^2 \sigma}}_{\alpha_{RP}}.
 \label{eqn:atten_water}
\end{equation}
The results indicate the existence of a fully broken ice limiting behaviour, in which the wave fields for different realisations of the ice cover are almost identical (Figure~\ref{Fig:Ensemble_Distro}g--i) and their properties are insensitive to further reduction of mean floe length.

\section{Homogenisation for Fully Broken Ice Regime}

\subsection{Bloch Waves in Periodic Ice Covers}\label{sec:bloch_1}

To determine the properties of wave fields in the fully broken regime 
(Figures~\ref{Fig:Ensemble_Distro} and~\ref{Fig:FloeLen_Mean}), a periodic ice cover composed of floes of constant length, $\bar{l}$, separated by zero open water gaps, $l_g=0$, is considered. 
The $j^{th}$ floe and surrounding open water (Figure~\ref{Fig:ModelSchematic}b) is taken as the representative unit cell of the periodic ice cover.
Without loss of generality, the left-hand edge of the floe is set as $x_j=0$ and, thus, the right-hand edge is $x=\bar{l}$, and, hence, subscripts $j$ are dropped for brevity.

Solutions are sought to \eqref{eqn:Gov_Eqn} in the unit cell, which satisfy Bloch's theorem, such that
 \begin{equation}
 \phi(x + \bar{l},z) = e^{i q \bar{l}}\phi(x,z) \quad \text{ and } \quad 
 \frac{\partial}{\partial x}\phi(x + \bar{l},z) = e^{i q \bar{l}} \frac{\partial}{\partial x}\phi(x,z),
\label{eqn:BlochCond}
\end{equation}
where $q\in\mathbb{C}$ is a Bloch wavenumber. 
The velocity potential is the open water at the left-hand edge of the floe, $\phi(0_{-},z)$, is defined by the amplitudes of the leftward  and rightward  wave modes,
$\mathbf{a}^{-}$ and $\mathbf{b}^{-}$, which both have length $M+2$ (two added `numerical' modes due to the very thin ice cover). 
Similarly, the velocity potential in the open water at the right-hand edge of the floe, $\phi(\bar{l}_{+},z)$, is defined by $\mathbf{a}^{+}$ and $\mathbf{b}^{+}$. 

The amplitudes in the open water at the right-hand edge of the floe are related to those at the left-hand edge using a transfer matrix, $\mathbf{P}$, like so
\begin{equation}
  \begin{bmatrix}
  \mathbf{a}^{+} \\
  \mathbf{b}^{+}
 \end{bmatrix} = \underbrace{\mathbf{V}_{R} \begin{bmatrix}
		\mathbf{\tilde{D}}^{a} & \mathbf{0} \\
		\mathbf{0} & \mathbf{\tilde{D}}^{b}
	\end{bmatrix} \mathbf{V}_{L}}_{\equiv\mathbf{P}} \begin{bmatrix}
  \mathbf{a}^{-} \\
  \mathbf{b}^{-}
 \end{bmatrix}.
 \label{eqn:Transfer_Cell}
\end{equation}
The matrix
\begin{equation}
 \mathbf{V}_{L} = \begin{bmatrix}
\mathbf{{T}}^{-} - \mathbf{{R}}^{+}\left(\mathbf{{T}}^{+} \right)^{-1}\mathbf{{R}}^{-} & \mathbf{{R}}^{+}
\left(\mathbf{{T}}^{+} \right)^{-1} \\
- \left(\mathbf{{T}}^{+} \right)^{-1} \mathbf{{R}}^{-} & \left(\mathbf{{T}}^{+} \right)^{-1}
\end{bmatrix}
\end{equation}
describes the scattering of wave modes at $x={0}$, relating the amplitudes in the open water to those in the floe covered water. 
Using symmetry, the corresponding scattering matrix at $x=\bar{l}$, $\mathbf{V}_{R}$, is
\begin{equation}\label{eq:VR}
 \mathbf{V}_{R} = \begin{bmatrix}
		\mathbf{0} & \mathbf{I} \\
		\mathbf{I} & \mathbf{0} 
	\end{bmatrix} \mathbf{V}_{L} ^{-1}
	\begin{bmatrix}
		\mathbf{0} & \mathbf{I} \\
		\mathbf{I} & \mathbf{0} 
	\end{bmatrix}.
\end{equation}
The edge transfer matrices, $\mathbf{V}_{R}$ and $\mathbf{V}_{L}$, do not have full rank since the water's motion is not influenced by the additional numerical modes ($\kappa_{M+1}$, $\kappa_{M+2}$) and,  hence, the inverse on the right-hand side of \eqref{eq:VR} is a Moore-Penrose pseudo-inverse.
The diagonal matrix $\mathbf{\tilde{D}}^{a}$ in (\ref{eqn:Transfer_Cell}) describes the phase change and modulation of the rightward propagating/decaying wave modes in the floe covered water from $x={0}$ to those at $x=\bar{l}$, and $\mathbf{\tilde{D}}^{b} = (\mathbf{\tilde{D}}^{a})^{-1}$ is the corresponding matrix for the leftward  wave modes. 

Bloch's theoreom \eqref{eqn:BlochCond} implies that each of the $2(M+2)$ eigenvalues of $\mathbf{P}$ correspond to length $2(M+2)$ vector of Bloch wavenumbers \cite{Griffiths_2001} 
\begin{equation}
 \mathbf{q}= \frac{1}{i \bar{l}} \left[ \ln\left( \boldsymbol{\upsilon} (\mathbf{P})\right)\right],
 \label{eqn:Wavenumber_Sol}
\end{equation}
where $\boldsymbol{\upsilon}(\mathbf{P})$ is the vector of the eigenvalues of $\mathbf{P}$. 
The Bloch wavenumbers come in positive and negative pairs due to \eqref{eqn:Transfer_Cell} describing the transfer of leftward and rightward propagating wave modes. 

The $M+2$ positive wave modes correspond to rightward propagation/decay and contain one damped propagating wave mode (positive real and imaginary parts), $M-1$ evanescent modes (complex modes with positive real parts three orders of magnitude smaller than their positive imaginary parts) and two numerical modes (large positive real and imaginary parts, $>100$\,m$^{-1}$). 
The damped propagating wave mode is the Bloch wavenumber of interest and, thus, denoted $q_b$. 
The corresponding wavelength and attenuation rate are, respectively,
\begin{equation}
 \lambda_b = 2 \pi / \Re \left\lbrace q_b\right\rbrace
 \quad\text{and}\quad 
 \alpha_b = \Im\left\lbrace q_b\right\rbrace.
\end{equation}

The amplitude transmitted into the periodic ice cover is obtained by considering $N$ identical floes from $x=0$ to $x = \bar{l}N$ in the limit where $N\to\infty$. A unit amplitude, rightward propagating wave mode is incident on the ice cover from the open water on the left of the floes with no incident wave modes from the open water on the right, i.e., $\mathbf{a}^{-}_{1} = [1,0,\dots,0]$ and $\mathbf{b}^{+}_{N} = [0,\dots,0]$. The ice cover reflects leftward propagating wave modes into the open water occupying $x<0$ and transmits rightward propagating wave modes into the open water on the right occupying $x>\bar{l}N$, with respective amplitudes $\mathbf{b}^{-}_{1}$ and $\mathbf{a}^{+}_{N}$ that are to be determined. These amplitudes are related by a transfer matrix across $N$ floes, which is the matrix $\mathbf{P}^N$. 

The eigen-decomposition of $\mathbf{P}^N$ is
\begin{multline}
\mathbf{P}^N = \mathbf{W}^{-1}\boldsymbol{\Upsilon}^N\mathbf{W}
\enspace\text{where}\enspace
\boldsymbol{\Upsilon} = \begin{bmatrix}
		\mathbf{Q} & \mathbf{0} \\
		\mathbf{0} & \mathbf{Q}^{-1} \end{bmatrix}
 \enspace\text{and}\enspace 
\mathbf{Q}= \text{diag}\left\lbrace \left[ e^{i q_1 \bar{l}}, \dots, e^{i q_{M+2} \bar{l}}\right] \right\rbrace.
\label{eqn:EigenDecomp}
\end{multline} 
For $N$ floes, the wave modes either side of the ice cover are related by
\begin{equation}
 \begin{bmatrix}
		\mathbf{a}^{+}_{N} \\
		\mathbf{b}^{+}_{N}
	\end{bmatrix}
 =
 \begin{bmatrix}
		\left( \mathbf{W}^{-1}\right)_{1,1} & \left(\mathbf{W}^{-1}\right)_{1,2} \\
		\left(\mathbf{W}^{-1}\right)_{2,1} & \left( \mathbf{W}^{-1}\right)_{2,2}\end{bmatrix} \begin{bmatrix}
		\mathbf{Q}^N & \mathbf{0} \\
		\mathbf{0} & \mathbf{Q}^{-N}\end{bmatrix} 
 \begin{bmatrix}
		\mathbf{W}_{1,1} & \mathbf{W}_{1,2} \\
		\mathbf{W}_{2,1} & \mathbf{W}_{2,2}\end{bmatrix} 
 \begin{bmatrix}
		\mathbf{a}^{-}_{1} \\
		\mathbf{b}^{-}_{1}
	\end{bmatrix}.
	\label{eqn:SemiInfProblem}
\end{equation}
The bottom row of \eqref{eqn:SemiInfProblem} 
can be solved to obtain
\begin{multline}
 \label{eq:Rn}
	\mathbf{b}^{-}_{1} = \left(\left( \mathbf{W}^{-1}\right)_{2,1} \mathbf{Q} ^N \mathbf{W}_{1,2} + \left( \mathbf{W}^{-1}\right)_{2,2} \mathbf{Q}^{-N} \mathbf{W}_{2,2} \right)^{-1} \\ \left(-\left( \mathbf{W}^{-1}\right)_{2,1}\mathbf{Q}^N \mathbf{W}_{1,1} - \left( \mathbf{W}^{-1}\right)_{2,2} \mathbf{Q} ^{-N}\mathbf{W}_{2,1}\right)\mathbf{a}^{-}_{1} .
\end{multline}
The reflected amplitudes for an infinite number of floes, $\mathbf{r}_{\infty}$, is found by letting $N\to\infty$ in \eqref{eq:Rn}, to give
\begin{equation}
	\mathbf{r}_{\infty} = \lim_{N\rightarrow \infty} \mathbf{b}^{-}_{1} = -\left(\mathbf{W}_{2,2} \right)^{-1} \mathbf{W}_{2,1}\mathbf{a}^{-}_{1}.
 \label{eqn:Rinf}
\end{equation} 
For an infinite number of floes, since $\mathbf{a}^{-}_{1}$ and $\mathbf{b}^{-}_{1} =\mathbf{r}_{\infty}$ are known, the amplitudes at $x=\bar{l}$ (in the open water between the first and second floes) are given by
\begin{equation}
  \begin{bmatrix}
  \mathbf{{a}}_{1}^{+} \\
  \mathbf{{b}}_{1}^{+}
 \end{bmatrix} = \mathbf{P} \begin{bmatrix}
		\mathbf{a}^{-}_{1} \\
		\mathbf{r}_{\infty}
	\end{bmatrix}.
 \label{eqn:Transmitted_Gap}
\end{equation}
The transferred amplitude into the periodic ice-cover, $A_b$, is approximated by the displacement generated in the first open water gap, given by
\begin{equation}
  A_b = \sum_{m=1}^{M} a_{m,1}^+ + b_{m,1}^+,
 \label{eqn:Transmitted_Amp}
\end{equation}
where the contributions of of the two numerical wave modes in the open water gaps are neglected. 

For the purposes of calculating $\lambda_b$, $\alpha_b$ and $A_b$ as $\bar{l}$ varies, 
a different formulation of the transfer matrix is used
(Appendix~\ref{App:TransMatrix}),  
which has better numerical stability than \eqref{eqn:Transfer_Cell}, as $\mathbf{V}_R$ and $\mathbf{V}_L$ do not have full rank. 
However, the transfer matrix formulation
\eqref{eqn:Transfer_Cell} facilitates analysis in the limit $\bar{l}\to{}0$ in \S\ref{sec:bloch_4}.

\subsection{Fully broken limits: $\bar{l} \rightarrow 0$}\label{sec:bloch_4}
The $\bar{l}$-dependence of $\mathbf{P}$ in (\ref{eqn:Transfer_Cell}) is contained in the diagonal matrices $\mathbf{\tilde{D}}^{a}$ and $\mathbf{\tilde{D}}^{b}$. 
Using the Taylor series expansion of the entries of the diagonal matrices, the right-hand side of (\ref{eqn:Transfer_Cell}) gives
\begin{align}
 &\mathbf{P} \sim \mathbf{I} + i\bar{l} \underbrace{ \mathbf{V}_R \begin{bmatrix}
\text{diag}\left\lbrace \mathbf{k} \right\rbrace& \mathbf{0} \\
\mathbf{0} & -\text{diag}\left\lbrace\mathbf{k} \right\rbrace
\end{bmatrix} \mathbf{V}_L}_{\equiv \,\mathbf{P}_1} 
\quad\text{for}\quad \bar{l}\ll 1, 
\label{eqn:P_TS_VrVl}
\end{align}
where $\mathbf{P}_{1}$ is independent of $\bar{l}$ and $\mathbf{V}_R\mathbf{V}_L = \mathbf{I}$ has been used.
Assume the regular limits
\begin{equation}
 \mathbf{W} \sim \mathbf{W}_{lim} + O(\bar{l})
 \quad\text{and}\quad
 \mathbf{q} \sim \mathbf{q}_{lim} + O(\bar{l})
 \quad\text{for}\quad \bar{l}\ll 1,
\end{equation}
so that $\mathbf{W} \to \mathbf{W}_{lim}$ and $\mathbf{q} \to \mathbf{q}_{lim}$ as $\bar{l}\to 0$,
then \eqref{eqn:EigenDecomp} gives an equivalent representation of $\mathbf{P}$ as
\begin{align}
&\mathbf{P} \sim \mathbf{I} + i\bar{l} \left(\mathbf{W}_{lim}\right)^{-1} \begin{bmatrix}
\text{diag}\left\lbrace \mathbf{q}_{lim} \right\rbrace & \mathbf{0} \\
\mathbf{0} & -\text{diag}\left\lbrace \mathbf{q}_{lim} \right\rbrace
\end{bmatrix}\mathbf{W}_{lim}
\quad\text{for}\quad \bar{l}\ll 1.
\label{eqn:P_TS_Eig}
\end{align}
Comparing \eqref{eqn:P_TS_VrVl} and \eqref{eqn:P_TS_Eig}, it can be deduced that
\begin{equation}
 \mathbf{P}_1 = \left(\mathbf{W}_{lim}\right)^{-1} \begin{bmatrix}
\text{diag}\left\lbrace \mathbf{q}_{lim} \right\rbrace & \mathbf{0} \\
\mathbf{0} & -\text{diag}\left\lbrace \mathbf{q}_{lim} \right\rbrace
\end{bmatrix}\mathbf{W}_{lim}. 
\end{equation}
Therefore, the fully broken limit of the Bloch wavenumbers is
\begin{equation}
 \mathbf{q}_{lim} = \boldsymbol{\upsilon}\left( \mathbf{P}_1\right),
 \label{eqn:WavenumberLim_Sol} 
\end{equation}
where $\boldsymbol{\upsilon}(\mathbf{P}_1)$ is the vector of the eigenvalues of $\mathbf{P}_1$. 
Only one of the limiting Bloch wavenumbers is a suitable damped, rightward propagating wavenumber, which is denoted $q_{lim}$. Note that the apparent diagonalisation of $\mathbf{P}_1$ by $\mathbf{V}_R$ and $\mathbf{V}_L$ ($\mathbf{V}_R\mathbf{V}_L = \mathbf{I}$) is distinct from $\mathbf{W}_{lim}$ since $\mathbf{V}_R$ and $\mathbf{V}_L$ do not have full rank.

The limiting transmitted amplitude, $A_{lim}$, is given by the limiting transfer matrix for $N$ floes, such that 
\begin{subequations}\label{eqn:EigenDecomp_Lim}
\begin{equation}
\mathbf{P}_{lim}^N = \mathbf{W}_{lim}^{-1}\boldsymbol{\Upsilon}^N\mathbf{W}_{lim}
\end{equation}
where
\begin{equation}
\boldsymbol{\Upsilon} = \begin{bmatrix}
		\mathbf{Q}_{lim} & \mathbf{0} \\
		\mathbf{0} & \mathbf{Q}_{lim}^{-1} \end{bmatrix}
  \\
\enspace\text{and}\enspace\mathbf{Q}_{lim}= \text{diag}\left\lbrace \left[ e^{i q_{lim,1} \bar{l}}, \dots, e^{i q_{lim,M+2} \bar{l} }\right] \right\rbrace.
\end{equation}
\end{subequations}
All of the following calculations follow as in \S\ref{sec:bloch_1} to give $A_{lim}$ as with $A_{b}$ in \eqref{eqn:Transmitted_Amp}.

\subsection{Results}
\label{Sec:Res_1s}

\begin{figure}
\includegraphics[width=\textwidth]{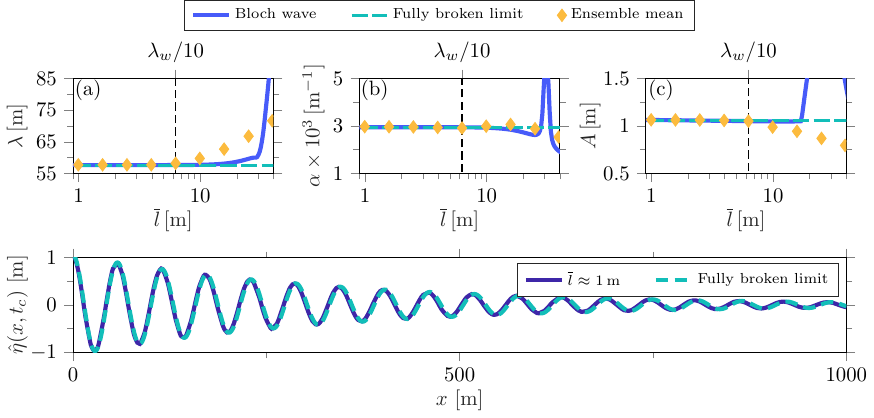} 
\caption{(a,b,c)~Comparison of Bloch waves with the ensemble means of (a)~$\lambda$, (b)~$\alpha$ and (c)~$A$, versus mean floe length. (d)~Wave field for the fully broken limit and the numerical solution for  ensemble member with $\bar{l} \approx 1 \, \text{m}$. }
\label{Fig:Results_1s}
\end{figure}

Figure~\ref{Fig:Results_1s}a--c shows the Bloch wave solutions for the wavelength, attenuation rate and transferred amplitude (\S{}\ref{sec:bloch_1})
and their fully broken limits (\S{}\ref{sec:bloch_4}) overlaid on the ensemble means (from Figure~\ref{Fig:FloeLen_Mean}) 
for floe lengths $\bar{l} < \lambda_w/2 $. 
The Bloch solutions accurately reproduce the wavelength, attenuation rate and transferred amplitude in the fully broken ice regime ($\bar{l} < \lambda_w/10\approx{}6.28\, \text{m}$), and the Bloch solutions are indistinguishable from their limits in this regime. 
(The agreement between the Bloch wave solution and ensemble means for attenuation also holds in the consolidated ice regime 
as the attenuation produced by a series of long floes is approximately the damping of consolidated ice 
since scattering is negligible.) 
Figure~\ref{Fig:Results_1s}d shows a comparison between a member of the ensemble in the fully broken regime (for $\bar{l} \approx 1 \,\text{m}$) and the fully broken limit for the wave field, i.e., using $q_{lim}$ \eqref{eqn:WavenumberLim_Sol} and $A_{lim}$. 
The example demonstrates the capability of the deterministic, semi-analytical Bloch wave approximation to reproduce the behaviour of realisations of the random ice covers in the fully broken regime.

\section{Dispersion Relations}

\begin{figure}
\centering
\includegraphics[width=\textwidth]{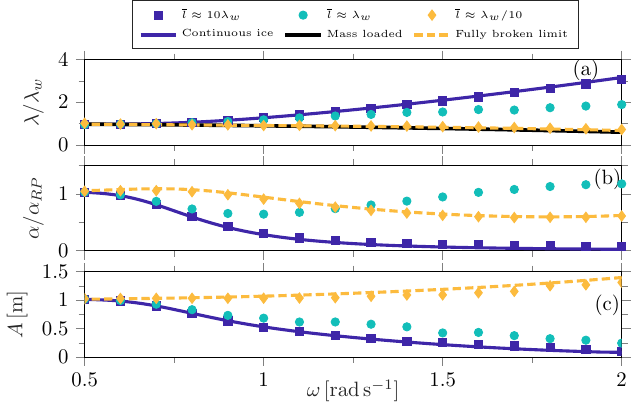} 
\caption{Dispersion relations for (a)~normalised wavelength, (b)~normalised attenuation rate and (c)~transferred amplitude, for ensemble means of numerical solutions (markers), continuous ice covers and the fully broken limit. }
\label{Fig:Results_MultPer}
\end{figure}

Dispersion relations in the consolidated ($\bar{l} \approx 10 \lambda_w$), transition ($\bar{l} \approx \lambda_w$) and fully broken ($\bar{l} \approx \lambda_w/10$) ice cover regimes 
are generated from ensembles of solutions over a range of angular frequencies, $\omega$, using $300$ randomly generated realisation for each ensemble. 
Figure~\ref{Fig:Results_MultPer} shows ensemble means of wavelength (normalised  by the open water wavelength),  attenuation (normalised by $\alpha_{RP}$)  and the transmitted amplitude, alongside the corresponding results given by a continuous ice cover (flexural-gravity wave dispersion), a continuous mass loaded ice cover ($F=\Gamma =0$; for the wavelength only) and the fully broken limit (\S{}\ref{sec:bloch_4}). 



For the normalised wavelength (Figure~\ref{Fig:Results_MultPer}a), when $\omega<0.8\,\text{rad}\,\text{s}^{-1}$, all ensemble means give wavelengths close to the open water wavelength, $\lambda_w \approx 2\pi / \sigma = 2\pi g/\omega^2 > 95$\,m (assuming deep water). 
As $\omega$ increases above $0.8\,\text{rad}\,\text{s}^{-1}$, the consolidated ice cover wavelength 
becomes increasingly greater than $\lambda_w$, in agreement with the continuous ice cover. 
In the transition ice cover regime, 
wavelengths also become increasingly greater than $\lambda_w$, but at a slower rate than the consolidated ice cover. 
Wavelengths in the fully broken regime are slightly shorter than open water wavelengths over the considered frequency range in agreement with a mass loaded ice cover. (Agreement between the fully broken limit and a continuous mass loaded ice cover was confirmed for a suite of numerical methods for both elastic and rigid floating plates \cite{Yiew-2019, Meylan-1994}.)


For the normalised attenuation rate (Figure~\ref{Fig:Results_MultPer}b), when $\omega < 0.6 \,\text{rad}\,\text{s}^{-1}$ ($\lambda_w > 174\,$m) the attenuation rate of the ensembles are approximately $\alpha_{RP}$ \eqref{eqn:atten_water}, since the wavelengths in the ice covers are approximately $\lambda_w$. 
As $\omega$ increases, the normalised attenuation rate of the ensembles decrease, with the consolidated ice cover decreasing the most. 
The normalised attenuation rate of the transition regime exhibits a shift from being dissipation dominated for $\omega < 1 \, \text{rad}\,\text{s}^{-1}$ ($\lambda_w >128 \,$m)
where it agrees with the consolidated ice cover, to being scattering dominated,
where the normalised attenuation increases with frequency, such that it is the ice cover with the largest attenuation rate of all the regimes when $\omega > 1.4 \, \text{rad}\,\text{s}^{-1}$ ($\lambda_w < 32 \,$m). 
The fully broken regime remains relatively close to $\alpha_{RP}$ for $\omega < 1 \,\text{rad}\,\text{s}^{-1}$ ($\lambda_w > 63\,$m) and then decreases to below $\alpha_{RP}$ for $\omega > 1.4 \,\text{rad}\,\text{s}^{-1}$ ($\lambda_w < 32\,$m) in agreement with the fully broken limit. 

For the range of parameters investigated (including frequency),  $\Gamma^2 \sigma \ll 1$, so that 
\begin{equation}
 \alpha_{RP} = \frac{ \sigma \sqrt{\sigma} \Gamma}{1 + \Gamma^2 \sigma} \approx \sigma \sqrt{\sigma} \Gamma = \omega^3 \frac{\Gamma}{ g \sqrt{g}}.
\end{equation}
Therefore, for frequency intervals where the normalised attenuation rate is insensitive to frequency, e.g., for the fully broken ice regime at high frequencies, the attenuation rate  is proportional to $\omega^3$, similar to observations \cite{Meylan-2018a}. 
The proportionality of $\alpha$ to $\omega^3$ for small $\omega$ is a direct feature of Robinson-Palmer damping, and is, thus, reproduced by the ensembles, continuous ice covers and fully broken limit. 
For the consolidated regime, the proportionality ceases as $\omega$ increases and $\alpha \rightarrow 0$. 
For the transition and fully broken regimes, the proportionality ceases and reemerges as $\omega$ increases, resulting in $\alpha$ being well described by some proportionality with $\omega^3$ for a large range of $\omega$ values for both regimes.

The transferred amplitude of all ensembles approach unity for $\omega < 0.6\,\text{rad}\,\text{s}^{-1}$ ($\lambda_w > 174\,$m), as the wavelengths in the ice covered water and the open water are similar, producing little scattering, resulting in no amplitude drop across the ice edge. As $\omega$ increases, the ice edge amplitude drop increases in the consolidated and transition regimes, as the waves supported in these ice covers are significantly longer than in open water and, thus, there is significant scattering. The amplitude drop for the consolidated regime agrees with the results for a continuous ice cover. The fully broken regime has an negative amplitude drop, in agreement with the fully broken limit. The negative amplitude drop is due to the wavelength in the fully broken ice cover being shorter than the open water wavelength. 

\section{Conclusions and Discussion}

A theoretical model was developed to study the effect of mean floe length (simulating the degree of ice breakup) on the propagation of waves from open water into and through high concentration ice covers. 
It was shown that ice covers with mean floe lengths greater than approximately ten open water wavelengths behave as consolidated ice covers, for which incident waves experience a large amplitude drop at the ice edge, become longer in the ice covered region and attenuate over distance.
Further, it was shown that there is a deterministic limiting behaviour for fully broken ice covers, attained when mean floe lengths are less than approximately one-tenth of the open water wavelength.
In the fully broken limit, there is a negative amplitude drop, waves become slightly shorter than in the open water (mass loaded system), and the attenuation rate is greater than the attenuation rate for the corresponding consolidated ice cover.
The fully broken limit was modelled by Bloch waves in a periodic ice cover, which facilitated semi-analytical expressions in the limit as the mean floe length tends to zero. 

As the ice cover transitions from consolidated to fully broken, i.e., mean floe lengths are comparable to open water wavelengths, multiple wave scattering is a dominant process and the wave field depends on the particular realisation of the ice cover.
Spectral analysis and ensemble averaging were used to extract wavelengths, attenuation rates and transferred amplitudes in the ice covered region, which join the consolidated and fully broken limits for wavelengths, attenuation rates and transferred amplitudes with respect to mean floe lengths.
Thus, the change in distance waves penetrate into an ice covered region before and after the ice cover breaks up depends on the relative degree of change in the competing processes of ice edge amplitude drop (which decreases as the ice cover breaks) and attenuation (which increases).

The model and findings presented do not contradict the observations of Collins~et~al.~\cite{Collins-2015} and Ardhuin~et~al.~\cite{ardhuin2020ice}.
They merely highlight the likely role of the ice edge amplitude drop, and question the interpretation that the attenuation coefficient decreases following break up of a consolidated ice cover.
In contrast, the findings are in qualitative agreement with those from physical models \cite{Dolatshah-2018,Passerotti-2022}.
A quantitative comparison with the physical models in a future study will demand a time-dependent model that incorporates a wave-induced ice breakup criterion and subsequent evolution of the floe size distribution, both of which are under debate
\cite{Passerotti-2022,montiel2022theoretical}, and incorporation of other physical processes, such as ice edge compaction due to radiation stress \cite{williams2017wave}.

Wave dissipation models based on anelastic flexure of the ice cover are often used, largely motivated by the interpretation of greater attenuation in consolidated ice covers
\cite{ardhuin2020ice}.
Incorporation of anelastic flexural damping into our theory, in place of Robinson--Palmer damping, would cause the attenuation coefficient to reduce (rather than increase) in the transition from consolidated to broken ice covers.
This does not necessarily imply that anelastic flexure is not the dominant damping during the consolidated phase of the ice cover in the physical models (and in the field), as other (unmodelled) dissipative processes are likely to become active after the ice cover breaks, particularly floe--floe collisions \cite{Bennetts-2015,Yiew-2017,Herman-2019}
and wave overwash of floes \cite{Skene-2015,Skene-2021}. 
The fully broken limit was consistent for both elastic and rigid plates, as expected. Therefore, models of very small rigid plates \cite{Iida-2023} provide a good basis for incorporation and assessment of different dissipation mechanisms in this floe size regime. 

Dumas-Lefebvre and Dumont~\cite{dumas2023aerial} recently found that phase speeds observed in a consolidated ice cover were closer to those predicted by mass-loading dispersion than flexural-gravity dispersion.
Their findings contrast with the those of the present study where wavelengths in the consolidated ice cover are consistent with flexural-gravity dispersion, and transition to mass-loaded dispersion in the fully broken limit.
Our findings are intuitive in the context of the model, i.e., the consolidated and fully broken regimes ($l\gg{}\lambda_w$ and $l\ll{}\lambda_w$) are similar to the assumptions of flexural-gravity and mass-loaded dispersion, respectively.
It is possible that the consolidated ice cover observed by Dumas-Lefebvre and Dumont~\cite{dumas2023aerial} is a refrozen ice cover consisting of a collection of small floes bonded together by thin ice, with snow cover obscuring this structure.


In conclusion, the present study helps to understand and quantify the findings from physical models and field observations on transitions in wave propagation characteristics through the ice covered ocean as the ice cover transitions from consolidated to fully broken. 
In particular, the study highlights the competing roles of the ice edge amplitude drop and wave attenuation, and motivates incorporation of the former in future studies and wave/sea ice models.
The semi-analytical expressions derived for the key quantities in the consolidated and fully broken limits provide a platform for future model developments.










\enlargethispage{20pt}




\appendix
\section{Maxima Method for Wave Property Quantification}
\label{App:Maxima}
\begin{figure}[h!]
\centering
\includegraphics[width=\textwidth]{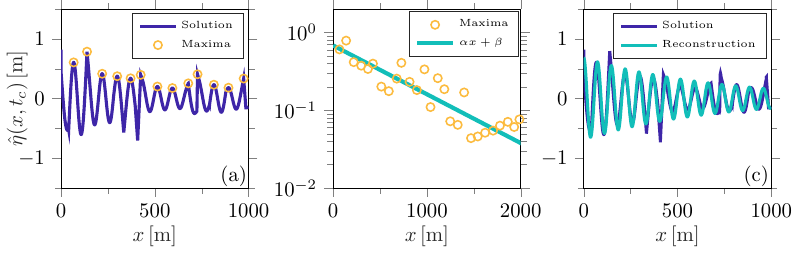} 
\caption{(a) The displacement of the ice cover with $\bar{l} \approx \lambda_w$ with maxima highlighted. (b) The log-lin plot of the maxima values compared to the fitted regression ($\alpha x + \beta$). (c) Comparison of $\hat{\eta}(x,t_c)$ and the reconstruction based on the measured wavelength, attenuation rate and amplitude. }
\label{Fig:Maxima_Method}
\end{figure}

For the maxima method, the displacement $\hat{\eta}(x,t_c)$ over the whole domain is taken. The local maxima of the displacement are identified, producing their locations $p_{1},p_{2},\dots,p_{O}$ and corresponding values $ \hat{\eta}(p_1,t_c), \hat{\eta}(p_2,t_c),\dots,\hat{\eta}(p_O,t_c)$ (Figure~\ref{Fig:Maxima_Method}a). 
The average distance between successive maxima locations provides an estimate of the overall wavelength and, thus, for the maxima method
\begin{equation}
 \lambda = \frac{1}{O-1}\sum_{o=1}^{O-1} \left| p_{o+1} - p_{o} \right|.
\end{equation}

As demonstrated in Figure~\ref{Fig:Maxima_Method}b there is a log-linear trend in the maxima values due to scattering and dissipation attenuating wave amplitude. To describe the trend in maxima values, the log-linear regression
\begin{equation}
 \ln(\hat{\eta}(p_o,t_c)) \sim \alpha p_o + \beta
\end{equation}
is fitted, which produces the attenuation rate $\alpha$ as the slope, and $\beta$ as the intercept. The transferred amplitude into the ice cover can be approximated by the fitted maxima value at the ice edge ($x=0$) so that $A = e^\beta $. The produced $\lambda$, $\alpha$ and $A$ using the maxima method result in a reasonable reconstruction of the displacement (Figure~\ref{Fig:Maxima_Method}c).

\section{Alternative Expression for Transfer Matrix}
\label{App:TransMatrix}
The scattering effect of the $j^{th}$ floe relating the amplitudes in the open water at $x=0$ and $x=\bar{l}$ (\S{}\ref{sec:bloch_1}) can be described by reflection and transmission matrices for the entire floe like so
\begin{align}
 \mathbf{a}^{+}= \mathbf{\widetilde{T}}^{-} \mathbf{a}^{-} + \mathbf{\widetilde{R}}^{+} \mathbf{b}^{+} \quad\text{and}\quad
 \mathbf{b}^{-} = \mathbf{\widetilde{R}}^- \mathbf{a}^{-} + \mathbf{\widetilde{T}}^+ \mathbf{b}^{+} .
  \label{eqn:RefTrans_ScatFloe}
 \end{align}
The reflection and transmission matrices\cite{Bennetts-2007} are given by the reflection and transmission matrices (Figure~\ref{Fig:ModelSchematic}c) like so
\begin{align}
 \mathbf{\widetilde{T}}^{-}&=\mathbf{{T}}^{-}\mathbf{\tilde{D}}^{a} \mathbf{U}_{1,1} \;,& 
 \mathbf{\widetilde{R}}^{+}&=\mathbf{{R}}^{+} + \mathbf{{T}}^{-} \mathbf{\tilde{D}}^{a} \mathbf{U}_{1,2 }\;, \\
 \mathbf{\widetilde{R}}^{-}&=\mathbf{{R}}^{-} + \mathbf{{T}}^{+} \mathbf{\tilde{D}}^{a}\mathbf{U}_{2,1} \;,&
 \mathbf{\widetilde{{T}}}^{+}&=\mathbf{{T}}^{+} \mathbf{{\tilde{D}}}^{a} \mathbf{U}_{2,2} 
\end{align}
where
\begin{align}
\mathbf{U} = \begin{bmatrix}
\mathbf{U}_{1,1} & \mathbf{U}_{1,2} \\
\mathbf{U}_{2,1} & \mathbf{U}_{2,2}
\end{bmatrix} = \begin{bmatrix}
\mathbf{I} & -\mathbf{{R}}^{+} \mathbf{\tilde{D}}^{a} \\
 -\mathbf{{R}}^{-} \mathbf{\tilde{D}}^{a}& \mathbf{I} 
\end{bmatrix}^{-1}\begin{bmatrix}
\mathbf{{T}}^{-} & \mathbf{0} \\
\mathbf{0} & \mathbf{{T}}^{+}
\end{bmatrix}.
\end{align}
The scattering equations \eqref{eqn:RefTrans_ScatFloe} can be rearranged into the form \eqref{eqn:Transfer_Cell},
with the associated transfer matrix 
\begin{equation}
 \mathbf{P} = \begin{bmatrix}
\mathbf{\widetilde{T}}^{-} - \mathbf{\widetilde{R}}^{+} \left(\mathbf{\widetilde{T}^{+}} \right)^{-1} \mathbf{\widetilde{R}}^{-} & \mathbf{\widetilde{R}^{+}}\left(\mathbf{\widetilde{T}^{+}} \right)^{-1} \\
- \left(\mathbf{\widetilde{T}^{+}} \right)^{-1}\mathbf{\widetilde{R}^{+}}& \left(\mathbf{\widetilde{T}^{+}} \right)^{-1}
\end{bmatrix} .
\label{eqn:Floe_Transfer}
\end{equation}
The dependence of $ \mathbf{P} $ on $\bar{l}$ 
in this form is not straightforward to extract due to the form of $\mathbf{U}$. 

\bibliography{Bib.bib}  

\end{document}

%% file: arxiv-luke.tex
\usepackage{amsmath}

\input{definitions-luke.tex}
\input{annotations-luke.tex}
\input{tikz-luke.tex}
\input{colours-luke.tex}

%% file: definitions-luke.tex
\renewcommand{\exp}{{\rm exp}}

\renewcommand{\Re}{\mathrm{Re}\,}
\renewcommand{\Im}{\mathrm{Im}\,}

\usepackage{enumerate}

\usepackage{overpic,rotating}

\newcommand{\beq}{\begin{equation}}
\newcommand{\eeq}{\end{equation}}
\newcommand{\bea}{\begin{eqnarray}}
\newcommand{\eea}{\end{eqnarray}}
\newcommand{\bean}{\begin{eqnarray*}}
\newcommand{\eean}{\end{eqnarray*}}

\newlength{\mylinelength}
\newlength{\mydashlength}
\newlength{\mydashspace}
\newlength{\mychainlengthA}
\newlength{\mychainlengthB}
\newlength{\mychainspace}
\newlength{\mylinethickness}
\usepackage{pifont}
\ifdefined\showmylabels
 \usepackage[inline]{showlabels}
 
\fi

\usepackage[normalem]{ulem}
\usepackage{soul}
\usepackage{tcolorbox}

%% file: annotations-luke.tex
\ifdefined\shownewold
   \newcommand{\del}[1]{{\color{red}\sout{#1}}}
\else
   \newcommand{\del}[1]{\ignorespaces}
\fi
\ifdefined\shownewold
   \newcommand{\deleqn}[1]{\\\del{\parbox{\textwidth}{#1}}}
\else
   \newcommand{\deleqn}[1]{\ignorespaces}
\fi
\ifdefined\shownewold
   \newcommand{\new}[1]{{\color[rgb]{0,0,1}#1}}
\else
   \newcommand{\new}[1]{#1}
\fi
\ifdefined\shownewold
   \newcommand{\old}[1]{{\color[rgb]{0.5,0.5,0.5}#1}}
\else
   \newcommand{\old}[1]{\ignorespaces}
\fi
\ifdefined\shownewold
   \newcommand{\delref}[1]{\del{\parbox{\figwidth\columnwidth}{#1}}}
\else
   \newcommand{\delref}[1]{\vspace{-12pt}}
\fi
\ifdefined\shownewold
   \newcommand{\lu}[1]{{\color[rgb]{1,0,1} #1}}
   \newcommand{\luchange}[2]{\del{#1}\new{#2}}
   \newcommand{\luleft}[1]{{\color[rgb]{1,0,1}$\longleftarrow$#1}}
   \newcommand{\luright}[1]{{\color[rgb]{1,0,1}#1$\longrightarrow$}}
\else
   \newcommand{\lu}[1]{\ignorespaces}
   \newcommand{\luchange}[1]{\ignorespaces}
   \newcommand{\luleft}[1]{\ignorespaces}
   \newcommand{\luright}[1]{\ignorespaces}
\fi


%% file: tikz-luke.tex
\usepackage{tikz,pgfplots}
\usetikzlibrary{calc,patterns,decorations.pathmorphing,decorations.markings,decorations.pathmorphing,angles,quotes,arrows,arrows.meta}
\tikzset{snake it/.style={decorate, decoration=snake}}

\newlength{\figwidth}

%% file: colours-luke.tex
\usepackage{xcolor}
\definecolor{darkgreen}{rgb}{0,0.55,0}
\definecolor{midgreen}{rgb}{0,0.8,0.2}
\definecolor{magenta}{rgb}{1,0,1}
\definecolor{purple}{rgb}{0.5,0,0.5}
\definecolor{darkorange}{rgb}{1,0.55,0}
\definecolor{maroon}{rgb}{0.5,0,0}
\definecolor{olive}{rgb}{0.5,0.5,0}
\definecolor{midgrey}{rgb}{0.5,0.5,0.5}
\definecolor{lightgrey}{rgb}{0.75,0.75,0.75}
\definecolor{matlabblue}{rgb}{0,0.447,0.741}
\definecolor{matlabred}{rgb}{0.85,0.325,0.098}
\definecolor{lightblue}{rgb}{0,0.5,1}
\definecolor{darkgrey}{rgb}{0.25,0.25,0.25}
\definecolor{teal}{rgb}{0,0.5,0.5}
\definecolor{navy}{rgb}{0,0,0.5}
\definecolor{goldenrod}{rgb}{0.85,0.6,0.1}